\documentclass{jaa}
\usepackage{natbib}
\usepackage{xcolor}

\usepackage{graphicx}
\usepackage{placeins}
\usepackage{float}

\begin{document}
\sloppy
\title{Examination of the Relationship Between Spectral Type and 
Stellar Rotational Velocity in $\sim$50,000 Single Stars}


\author{Boran Mert\textsuperscript{1}, Usta Ahmet\textsuperscript{1} and Kayhan 
Cenk\textsuperscript{1,2,*}}
\affilOne{\textsuperscript{1}Department of Astronomy and Space Sciences, 
Science Faculty, Erciyes University, 38030 Melikgazi, Kayseri, Türkiye}

\affilTwo{\textsuperscript{2}Scientific Research Projects Coordination Unit,
Kayseri University, 38280 Talas, Kayseri, Türkiye}


\twocolumn[{

\maketitle

\corres{cenkkayhan@erciyes.edu.tr}

\msinfo{ XX April 2024}{XX November 2024}

\begin{abstract}
In this study, we present the results of the relationship between spectral type (ST) and 
the projected stellar rotational velocity ($vsini$), utilising a sample of approximately 50,000 single stars across a range of 
evolutionary stages. The STs of the stars included in this study
span a broad range, from O0 to M9. 
We examine the stars in our data set, which has been divided into two
groups according to ST and luminosity class (LC). 
The groups were
conducted an investigation into the relationship between 
the mean $vsini$ ($\langle vsini \rangle$) and STs,
as well as the dependence of $\langle vsini \rangle$ on STs and LCs.
The rationale for investigating the two subgroups separately is
to take into account for the evolutionary status of the stars
and ascertain the impact on stellar rotation.
The results demonstrate a notable decline in
$\langle vsini \rangle$ as the spectral 
type progresses from early to late types. In particular, 
we found a significant decrease in $\langle vsini \rangle$ values, 
amounting to approximately 100 km/s, 
between hot stars (STs O0 to F2) and cool stars (STs F2 to 
M9). Moreover, a reduction in $\langle vsini \rangle$ is discernible as stars evolve, 
with this trend being most pronounced in 
evolutionary stages beyond the subgiant phase.
\end{abstract}

}]


\doinum{12.3456/s78910-011-012-3}
\artcitid{\#\#\#\#}
\volnum{000}
\year{2024}
\pgrange{1--}
\setcounter{page}{1}
\lp{1}

\section{Introduction}
All stars are born within molecular clouds, evolve over time, and ultimately reach the end of their evolution. 
The parameters that determine the manner in which this 
radiant life will unfold represent a complex puzzle. It is essential to gain an understanding of each piece and to place it correctly. Among these parameters, stellar rotation is a factor 
that exerts a significant influence on the overall structure of the star, from the stellar core to the stellar surface.
The most significant study on stellar rotation was conducted by 
\citet{1924MNRAS..84..665V}, which demonstrated that stellar rotation is a crucial parameter in stellar 
evolution. 
One of the most readily apparent consequences of stellar rotation is the alternation of stellar shape, manifested as a departure from spherical
symmetry \citep{2005ApJ...628..439M, 2011ApJ...732...68C}. Consequently, based on certain observations, there is the potential for a variation of 
up to 50\% between the radius of the polar and equatorial 
regions \citep{2009pfer.book.....M, 2023A&A...669L..11A}. Such alternations in radius result in a modified effective temperature profile 
and ionisation equilibrium of the star 
\citep{1924MNRAS..84..665V, 2009pfer.book.....M}.
An additional consequence of stellar rotation is the alternation of the chemical composition of the star.
Differential rotation disrupts the equilibrium of the stellar core and  layers, prompting the mixing of matter within the core and 
radiative envelope \citep{2018MNRAS.480.5427J, 2019MNRAS.485.4641C}. 
Some studies have demonstrated that rotational mixing can enhance the
 supply of hydrogen within the stellar core, thereby prolonging the lifespan of the 
main-sequence (MS) phase
by approximately 30\% \citep{2000A&A...361..101M}. 
In the initial stages (first few million years) of stellar evolution, as evidenced by multiple  studies (e.g. \citet{2006ApJ...646..297R, 
2007ApJ...671..605C, 2013A&A...556A..36G}), 
stars with accretion discs tend to exhibit slower rotation rates than those without discs.
The primary rationale for this phenomenon can be attributed to the disc-locking hypothesis, which posits that the star remains attached to the disc, thereby maintaining a constant 
projected stellar rotational velocity ($vsini$).
It has been demonstrated by various studies 
(e. g. \citet{1991ApJ...370L..39K, 1993A&A...274..309C, 1996MNRAS.280..458A}), 
that the interaction between a star and a stellar disc 
with a magnetic field can regulate $vsini$ of the central star,
resulting in the star becoming tidally locked \citep{2001ApJ...548.1071B}. This will 
result in a reduction in $vsini$ of the star.
However, when the star loses
its disc, it undergoes a period of rapid rotation as a consequence of the anticipated 
radius contraction along the Hayashi track \citep{2013A&A...556A..36G}.
As evidenced by prior studies in the literature, this phenomenon can be attributed to the inability of the star to sustain its angular momentum from the zero-age MS (ZAMS) to MS.

The magnetic effects exerted upon a star result in a reduction in its $vsini$, a process known as 'magnetic braking' \citep{1967ApJ...150..551K}. This is attributable 
to the magnetised stellar winds carrying angular momentum away from the star \citep{2016APJ...823...16B, 2024MNRAS.533.1290S}.
\citet{1972ApJ...171..565S} demonstrates 
that the surface dynamo area of a star diminishes in conjunction 
with both $vsini$ and time (or age) during the MS phase. 
\citet{1972Ap&SS..17..489D} corroborates this hypothesis with the aid of a stellar wind model. Late-type stars (G, K, M) experience magnetic braking 
from dynamo fields, whereas early-type stars (O, B, A, early-F) exhibit weaker fields due to the presence of less convection.
Conversely, the evidence 
presented by \citet{1991ApJ...376..204M} indicates that stars may attain a state 
of 'magnetic saturation' if they rotate at a high rate, which would serve to reduce 
the loss of angular momentum.
Moreover, \citep{2017NatAs...1E.186D} put forth  a potential
explanation for the bimodal rotational distribution observed in binary stars. 
This explanation postulates that magnetic wind braking or tidal torques exerted
by a companion of the binary star could rapidly decelerate $vsini$, thereby 
facilitating a transition from the rapidly rotating 
track to the non-rotating track \citep{2020IAUS..351..228S}.

The relationship between the distribution of 
$vsini$ 
and spectral type (ST) has been the subject of investigation 
since the 1960s \citep{2009pfer.book.....M}. 
However, these studies have often been limited to specific ST or
luminosity class (LC) ranges \citep{1962AnAp...25...18S, 1968BAN....19..309V, 
1979RA......9...87S, 1997ApJ...487..365A, 2007A&A...463..671R, 
2014prpl.conf..433B, 2023A&A...669L..11A}. 
In this study, we conduct
out using a large sample of stars of various ST
observed in different surveys. The objective of this study is to elucidate the 
relationship between stellar rotation
and the ST and LC, which are indirectly related to the 
effects of the stellar structure and evolution. 

The data employed in this study
are presented in the following section (Section 2). In Section 3,
we provide a methodology for examining the relationship between $vsini$ and ST and LC.
We discuss the results of this study in Section 4. Finally, 
the conclusion of this study is summarised in Section 5. 

\section{Data Set}
We select $\sim$ 50,000 stars with varying evolutionary statuses from catalogues in the literature that are presented in the Table~\ref{table1}.
In order to ensure the reliability of the analysis,
we excluded intrinsic variable stars, binary stars and chemically peculiar stars from our study 
that are mentioned in the catalogues.
Following the aforementioned exclusions, the total number of the stars examined in this study is 48,639.

\begin{table}[htb]
\tabularfont
\caption{The references, the total number of stars and their spectral types (ST) 
employed in this study are presented in alphabetical order.}
\begin{tabular}{lcl}
\topline
References & Total Number & ST \\
           & of Stars &  \\
\midline
{\citet{2013AJ....145..132C}} & 6   & A, F    \\
{\citet{2014A&A...566A.130C}} & 165  & M       \\
{\citet{2002A&A...384..491C}} & 198  & F, G, K \\
{\citet{1999A&AS..139..433D}} & 1540 & F, G, K \\
{\citet{2004A&A...427..313D}} & 159  & F, G, K \\
{\citet{2002ApJ...578..943D}} & 134  & F, G, K \\
{\citet{2014A&A...561A.126D}} & 1588 & F, G, K \\
{\citet{2002A&A...395...97D}} & 232  & F, G, K \\
{\citet{2005ESASP.560..571G}} & 36,419 & O, B, A, F, \\
                              &      & G, K, M \\
{\citet{2022A&A...665A.150H}} & 285  & O       \\
{\citet{1997MNRAS.284..265H}} & 185  & O, B    \\
{\citet{2018A&A...614A..76J}} & 1012 & M \\
{\citet{2019ApJ...879..105L}} & 221  & K, M \\
{\citet{2010A&A...520A..15M}} & 147  & F, G, K, M \\
{\citet{2011A&A...532A..10M}} & 99  & G, K, M \\
{\citet{2023A&A...676A..85S}} & 145  & O, B     \\
{\citet{2020ApJ...888...82S}} & 9   & O       \\
{\citet{2014A&A...562A.135S}} & 199  & O, B    \\
{\citet{2015PhDT.......701S}} & 48   & M       \\
{\citet{2002A&A...381..105R}} & 525  & B, A, F \\
{\citet{2002A&A...393..897R}} & 2401 & B, A, F, \\
                              &      & G, K  \\
{\citet{2007A&A...463..671R}} & 1541 & B, A, F \\
{\citet{2004ApJ...601..979W}} & 143  & O, B, A, \\
                              &      & F, G, K \\
{\citet{2007AJ....133.1092W}} & 123  & B       \\
{\citet{2012A&A...537A.120Z}} & 2012 & B, A, F \\
\hline
\label{table1}
\end{tabular}
\end{table}

\begin{figure}[!t]
\centering
\includegraphics[width=\columnwidth]{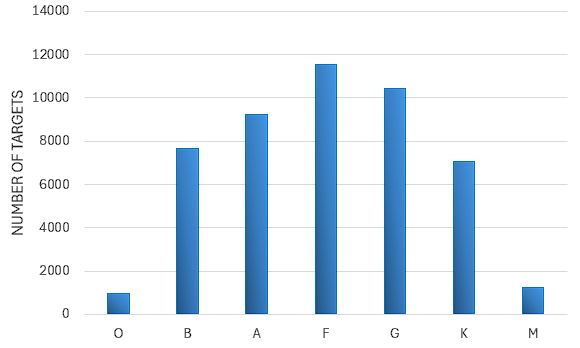}
\caption{The distribution of stars according to their STs in the data set is presented.}
\label{fighisto}
\end{figure}

The distribution of the stars in the data set
according to STs is presented as a histogram in Figure~\ref{fighisto}.
As is evident from the Figure~\ref{fighisto}, the majority 
of the data is concentrated in 
the F (11536) and G (10454) STs. The data for the B, A and K STs are
relatively similar, with numbers of 7670, 9249, and 7072, respectively.
The lowest recorded data belongs to stars belonging to the O (961) and M (1232) 
STs.

\begin{figure*}[!t]
\includegraphics[width=\textwidth]{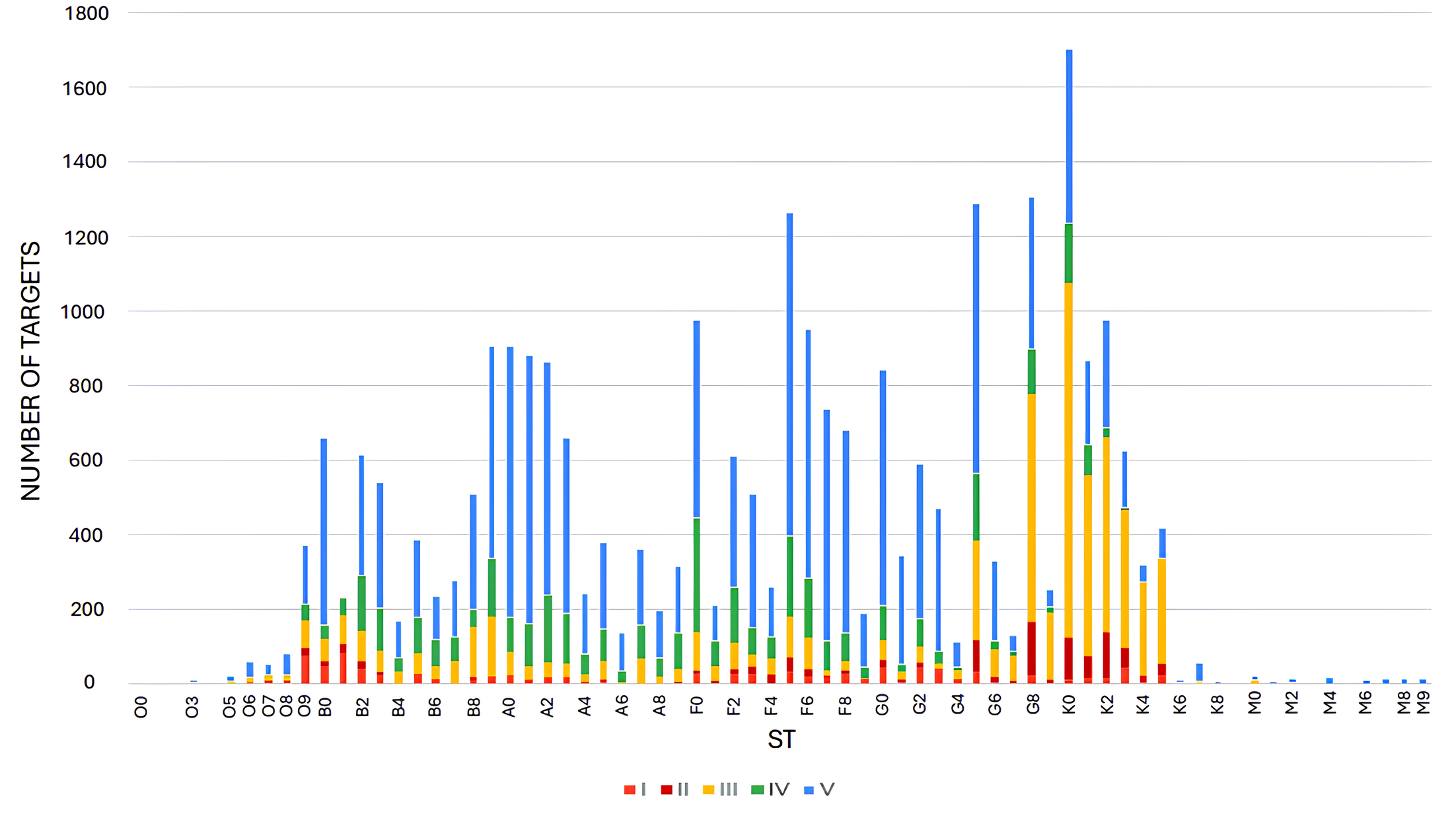}
\caption{The distribution of stars in the data set 
according to their detailed
STs 
and LCs. Each ST is 
divided into subclasses, as indicated by the numbers.
In the histogram the colours 
blue, green, yellow, red and orange 
represent the
MS (V), subgiant (IV), red giant (III), bright giant (II) 
and supergiant (I), respectively.}
\label{figpar2}
\end{figure*}

Additionally, the number distribution of LC for the stars is plotted in Figure~\ref{figpar2}.
The data set comprises five distinct LCs: MS (V), subgiant (IV), 
red giant (III), bright giant (II), and supergiant (I). As plotted in
Figure~\ref{figpar2}, the data set comprises the greatest number of 
MS stars and the fewest supergiant stars.

The general distribution of the number of stars for each ST is over 200.
The number of early-O stars is less than 200, due to the fact that  
these stars evolve rapidly and are difficult to detect. Similarly, the number of late-K
and M stars (as plotted in Figure~\ref{figpar2}) does not exceed 100. These stars are notoriously challenging to observe due to their low luminosity.

\section{Method}

A comprehensive methodology was employed to investigate 
the relationship between STs and $vsini$. The analysis 
commenced with the categorisation of the target stars 
into two distinct groups based on their STs and 
LCs.
The formation of these groups was undertaken to facilitate investigation into the relationship
between $vsini$-STs, as well as the relationship between $vsini$-STs in consideration of LCs.

The rationale for investigating
the two subgroups separately is
to account for the evolutionary status of the stars
and to determine the impact on
of this status on stellar rotation.

In order to facilitate the process of coding, 
each ST is assigned a numerical identifier in the study. 
Subsequently, the corresponding $vsini$ values 
for each ST were averaged. 
This enabled an examination of the $vsini$ values of the stars over a specified 
range (Fig.~\ref{figOne}). 
A Gaussian modelling was employed to derive
fitting variables for $\langle vsini \rangle$ and ST, which were found to 
follow a normal distribution. In order to achieve the optimal fit,
a model consisting of a three-component Gaussian
distribution was constructed. 
The model, based on a Gaussian fit,
utilises the amplitude ($a$), sigma ($b$) and centre ($c$) parameters as fitting 
variables (Eq.~\ref{eq:gx}). In order to 
constrain the model, full width at half maximum and maximum peak 
height parameters are also included. The amplitude, sigma and centre  
parameters
represent the strength of the peak, the characteristic width of the peak and centre value of the line to estimate sigma, respectively. The use of constrained 
parameters, such as
full width at half maximum and maximum peak height, allows for the comparison of models. 
The application of three Gaussian and decaying exponential
components in the models
facilitates the fitting of multiple peaks and the avoidance of  overfitting: 

\begin{equation}
\label{eq:gx}
g(x)=\sum_{i=1}^{3}{\frac{a_i}{\sqrt{2\pi b_i^2}}exp{{\left(-{\frac{(x-c_i)^2}{2b_i^2}}\right)}}}+a_iexp(-x/\tau)
\end{equation}

This approach eliminates the necessity
for
parameter boundaries in our models {\footnote{A code, designated
{\sc{pygaus}}, has been developed
in this study
which is capable of 
averaging and Gaussian modelling of the 
targets. This code can be
applied to Gaussian modelling across a range of ST.}}.

\section{Results and discussion}

\subsection{$\langle vsini \rangle$-ST relationship}

\begin{figure}[H]
\includegraphics[width=1\columnwidth]{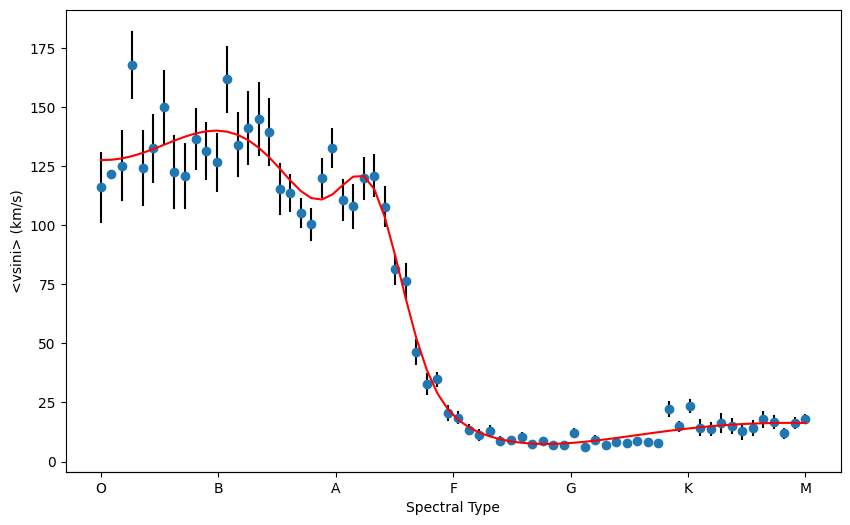}
\caption{The stellar rotation distribution of the stars in the data set, is represented by a plot of
$\langle vsini \rangle$ values 
on the y-axis and 
STs from O0 to M9 on the x-axis. The red line represents 
the Gaussian fit.}
\label{figOne}
\end{figure}

By employing the Gaussian modelling method, as previously outlined, 
we have established a general correlation (within the range of O0-M9 range) between
ST and $\langle vsini \rangle$, as illustrated in Figure~\ref{figOne}. Additionally, we have derived relations for the ten distinct 
STs, taking into account the underlying stellar
structure. The resulting 
fitting parameters for each
Gaussian component are presented in  
Table~\ref{tab:buyuk_tablo1}. 
In addition to Figure~\ref{figOne}, comprehensive plots for various ST ranges 
(O0-F0, F0-M9, O0-F1, F1-M9, O0-F2, F2-M9, O0-F3, F3-M9, O0-F4 and F4-M9), 
together with their corresponding fits, are presented in Appendix A. 

\begin{table}[!t]
\centering
\caption{The amplitude ($a$), sigma ($b$), and centre ($c$) 
parameters of the 
Gaussian components
obtained from the ith Gaussian fit for the different ranges of ST
are presented. To 
achieve the optimal fit, a model consisting of three Gaussians was 
constructed. The value of i represents the number of Gaussians used.}
\label{tab:buyuk_tablo1}
\begin{tabular}{ccccc}
\hline
& i & a & b & c   \\
\hline
\hline
& 1 & 104.15 & 9.47 & 16.39 \\
O0-M9 & 2 & 59.17 & 2.91 & 28.70 \\
& 3 & 15.60 & 10.21 & 65.09 \\
\hline
& 1 & 114.54 & 3.62 & 15.90 \\
O0-F0 & 2 & 118.63 & 6.31 & 27.42 \\
& 3 & 135.18 & 4.10 & 6.58 \\
\hline
& 1 & 9.30 & 12.69 & 31.73 \\
F0-M9 & 2 & 812.32 & 7.38 & 14.97 \\
& 3 & 16.29 & 9.92 & 65.06 \\
\hline
& 1 & 116.89 & 4.43 & 27.61 \\
O0-F1 & 2 & 132.30 & 4.20 & 16.53 \\
& 3 & 135.77 & 4.51 & 5.86 \\
\hline
& 1 & 47.03 & 0.04 & 61.37 \\
F1-M9 & 2 & 8.95 & 6.80 & 64.40 \\
& 3 & 338.71 & 5.44 & 22.01 \\
\hline
& 1 & 134.27 & 4.08 & 16.18 \\
O0-F2 & 2 & 117.72 & 4.74 & 27.47 \\
& 3 & 127.23 & 3.88 & 6.48 \\
\hline
& 1 & 11.44 & 7.33 & 35.61 \\
F2-M9 & 2 & 16.64 & 10.35 & 64.32 \\
& 3 & 40.00 & 2.90 & 31.33 \\
\hline
& 1 & 140.93 & 4.75 & 16.22 \\
O0-F3 & 2 & 113.96 & 4.14 & 27.87 \\
& 3 & 110.18 & 3.50 & 6.29 \\
\hline
& 1 & 16.60 & 9.87 & 64.53 \\
F3-M9 & 2 & 9.73 & 8.84 & 35.20 \\
& 3 & 33.91 & 3.29 & 31.39 \\
\hline
& 1 & 134.39 & 4.32 & 16.72 \\
O0-F4 & 2 & 116.80 & 4.13 & 27.72 \\
& 3 & 135.22 & 4.55 & 5.85 \\
\hline
& 1 & 153.23 & 4.91 & 25.58 \\
F4-M9 & 2 & 5.16 & 5.37 & 75.96 \\
& 3 & 10.05 & 7.68 & 63.74 \\
\hline
\hline
\end{tabular}
\end{table}

As illustrated in Figure~\ref{figOne}, the mean $vsini$ ($\langle vsini \rangle$) undergoes an abrupt decline 
for stars with effective temperatures below the F2 threshold. This phenomenon can be attributed to 
the displacement of the 
convective and 
radiative layers of the stars within
the stars, a phenomenon that has
been well-documented in the literature. $vsini$
of late-type stars is significantly reduced as a consequence of the magnetic brake 
produced by the dynamo process 
\citep{1967ApJ...150..551K, rozelot2009rotation, 2009MNRAS.392.1022U}. 

\citet{1972ApJ...171..565S} posits that the mean surface (dynamo) 
area is proportional to $vsini$ and decreases in proportion to
the inverse square root of time as the star undergoes nuclear fusion
on the MS. This remarkable result is demonstrated to be 
theoretically consistent on the basis of a simple model 
for the stellar wind proposed by \citep{1972Ap&SS..17..489D}.
The magnetic fields that cause magnetic braking in late-type stars (G, K and M)
are known to be generated by a dynamo mechanism in 
which the Coriolis forces associated with stellar rotation 
deflect convective motions in the recombination regions of H and He
\citep{2002ApJ...576..413U}. In contrast, early-type stars (O, B, A, and early-F) with higher initial angular momentum, 
shorter contraction timescales in the ZAMS, and fully ionised hydrogen in their atmospheres 
(lacking the strong convection regions associated with 
hydrogen recombination) are known to exhibit weaker 
dynamo-induced magnetic fields. This suggests 
that early-type stars (O, B, A, and early-F) 
are rotating at a faster rate than late-type stars (late-F, G, K, and M). 
Nevertheless, as the majority of hot stars 
typically rotate at high velocity, it is plausible that dynamo 
production in thin (weak) near-surface convection regions 
associated with the recombination of fully 
ionised helium may still occur \citep{2002ApJ...576..413U}.
In this regard, we found a significant decrease in $\langle vsini \rangle$ values was observed, 
amounting to approximately 100 km/s, between hot stars (spectral 
types O0 to F2) and cool stars (STs F2 to 
M9).

Figure~\ref{figOne} illustrates the high 
rotational velocities of O-type stars, which have been the subject of a 
recent study by \citet{2023AA...672A..22B}. This study examined and discussed
the potential relationship between 
these rapid rotation rates and binary star systems.
The findings revealed that 
a significant portion of the rapidly rotating O-type 
star population is associated with binary interactions.

The fastest stars in Figure~\ref{figOne}, 
as determined by their peak shapes, are of the B- and A-type stars.
Among them, the early-A stars, which manifest as a distinct peak
in Figure~\ref{figOne}, 
are the fastest stars. It has been observed that there is a notable decline 
in the velocity of late-A stars, with a reduction of approximately 40 km/s,
as they transition towards early-F stars.

In contrast to the convective envelopes observed in the late-A type stars,
those of early-A type stars are notably absent
\citep{1995APJS...99..135A, 2007A&A...463..671R}. 
Late-A type stars are situated in close proximity to the boundary 
at which the convective envelope 
becomes apparent in the stellar photosphere, as illustrated 
in the Hertzsprung-Russell diagram (HRD). 
It has been observed that late-A type stars are situated in close
proximity to the granulation boundary, a phenomenon that is exclusive to MS stars.
In the case of evolved late-A type stars, their position in the HRD
is found to be significantly bluer than the granulation boundary. 
A detailed discussion of this situation can be found in Section 4.2.
An additional factor contributing to this phenomenon is that examining 
stellar rotation as a mean can be somewhat deceptive. This is due to
the fact that early-A type stars exhibit both high and low 
stellar rotation velocities, which are consistent with observations presented by 
\citet{1995APJS...99..135A}. They identified a bimodal distribution 
of stellar rotational velocities among A-type MS stars.

A comparable phenomenon is observed in M-type stars, albeit with a considerably 
diminished rate of change in the $\langle vsini \rangle$, 
estimated at $\sim$20 km/s. As illustrated in Figure~\ref{figOne}, early-M
stars can be identified as a peak. 
\citet{1998AAA...331..581D} and \citet{2003csss...12..683M} 
posit that saturation-type rotation-activity is observed
in M5- to M8.5-type stars, resulting in an increase of approximately
5-10 km/s in the $vsini$ of these stars. 
As illustrated in Figure~\ref{figOne},
the $vsini$ of M9-type stars exhibits a notable decline due to  
a precipitous decline in activity \citep{2013AJ....146..156D}.
Moreover, metal-poor M-type stars are observed to 
be more compact and exhibit higher rotational velocities 
for a given mass \citep{2003csss...12..683M}. 
These effects may provide an explanation for the slight increase
in stellar rotation velocity observed in M-type stars. 
Figure~\ref{figOne} illustrates that late-M stars exhibit
a markedly higher $\langle vsini \rangle$
than their early-M counterparts. 
This may be attributed to the two dynamo processes that
are strongly coupled to the rotation in stars composed 
entirely of the convective layer \citep{2021LRSP...18....5F}.
It is important to note that the studies mentioned above only concern M dwarfs,
whereas our study also encompasses evolved M-type stars. 
We demonstrated that the overall $vsini$-ST distribution 
of our data, as represented by the Gaussian-fitted line in Figure~\ref{genel} 
of the Appendix, is consisted with the aforementioned findings.

\subsection{$\langle vsini \rangle$-LC relationship}

\begin{figure*}[!t]
\includegraphics[width=\textwidth]{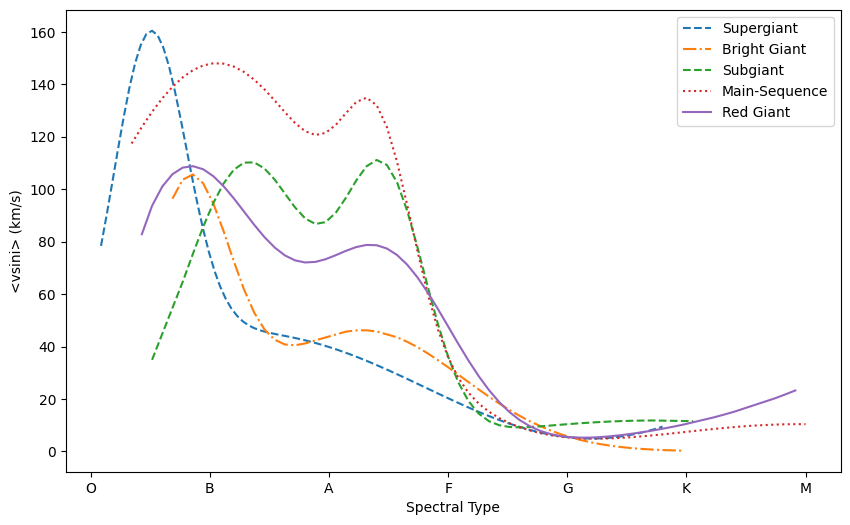}
\caption{
The distribution of $\langle vsini \rangle$ versus ST for single stars
of varying evolutionary statuses.
The x-axis and y-axis represent the STs from O0 to M9 and  
the $\langle vsini \rangle$ values, respectively. The Gaussian-fitted lines for 
supergiant, bright giant, red giant, subgiant and MS are represented by 
blue, orange, purple, green and red colours, respectively.}
\label{lc}
\end{figure*}

As previously stated in Section 2, our data set also encompasses stars 
that are at disparate evolutionary stages. 
Accordingly, the LCs were divided into five subgroups:
supergiants, bright giants, red giants, subgiants and MS stars. 
To elucidate the correlation between LC and $\langle vsini \rangle$,
the methodology employed for ST was replicated. 
The derived parameters for each LC are presented in Table~\ref{tab:buyuk_tablo2}.
Figure~\ref{lc} illustrates the distribution of $\langle vsini \rangle$ 
with ST for single stars at disparate evolutionary stages.

\begin{table}[!t]
\centering
\caption{The same methodology as that employed in Table~\ref{tab:buyuk_tablo1} 
is used here, but the LCs are divided into five subgroups, as follows:
supergiants, bright giants, 
red giants, subgiants and MS stars.}
\label{tab:buyuk_tablo2}
\begin{tabular}{ccccc}
\hline
& i & a & b & c   \\
\hline
\hline   
& 1 & 140.39 & 5.60 &  3.67   \\
Supergiant & 2 & 42.54 & 16.97 & 9.30    \\
& 3 & 10.72 & 36.58 & 14.08   \\
\hline
& 1 & 95.01  & 9.53 & 4.08   \\
Bright Gaint& 2 & 46.21 & 26.59 & 9.85    \\
& 3 & 161.31 & 79.47 & 2.79 \\
\hline
& 1 & 140.20  & 122.01 & 27.97   \\
Red Giant& 2 & 72.55 & 28.69 & 6.62    \\
& 3 & 108.16 & 9.34 & 7.75 \\
\hline
& 1 & 109.28  & 15.38 & 6.19   \\
Subgiant & 2 & 95.99 & 28.80 & 4.04    \\
& 3 & 11.82 & 55.37 & 17.17 \\
\hline
& 1 & 148.05  & 12.36 & 12.27   \\
MS & 2 & 66.36 & 28.21 & 3.35    \\
& 3 & 10.41 & 69.36 & 13.24 \\
\hline
\hline
\end{tabular}
\end{table}

It is a well-established fact that as stars evolve, 
their $vsini$ tends to decrease
\citep{1953PASP...65..192H, 1955APJ...121..118H, 
1965APJ...141..828K, 1985APJ...298..756G, 1986APJ...310..277G, 
1987APJ...322..360G}.
Figure~\ref{lc} illustrates the decline in $vsini$ 
that occurs during the later stages of the stellar evolution for each ST.
As the star exhausts its hydrogen in its core, radiation pressure cannot resist 
the force of gravity. As a result, the core undergoes a reduction in size,
yet the expansion of the hydrogen-burning shell within the envelope results in the
 the opposite effect. This leads to an increase in the moment
of inertia \citep{1967APJ...148..217W}. Consequently, as the star evolves towards 
the giant branch, the moment of inertia increases, which results in a 
decrease in stellar rotation \citep{1996AAA...314..499D, 1997APJ...480..303K}.
Given the increase in stellar radius that occurs 
during the later stages of MS evolution, 
a decrease in $vsini$ is to be expected. Moreover, 
mass loss during the subsequent evolutionary stages of the star also contributes
to this decrease. 
In the case of cool stars, the dynamo process and magnetic braking 
represent additional factors that contribute to the slowdown
of $vsini$
\citep{1981APJ...251..155G, 1982JRASC..76..319G, 1983IAUS..102..461G}.
In contrast, for hot stars, 
an additional contribution to the slowdown is provided by
stellar winds \citep{2022EL....14029001D}.
A decrease in $vsini$ 
is also observed in the MS for A-type evolved stars.
However, as stars evolve, the granulation boundary shifts to the hotter region 
in Figure~\ref{lc}, and the change is manifested in the hotter region
in accordance with the stages of evolution.
This situation is most obvious in the evolutionary stages after the subgiant 
evolutionary stage.

\section{Conclusion}

The STs of stars provide insight into their internal structure. The stellar
structure is influenced by various parameters, including $vsini$. 
In this study, we analysed the relation between 
$vsini$ and ST 
using data on approximately 50,000 non-variable and single stars 
at different evolutionary stages and with a broad range
range of STs from O0 to M9. 
Our findings revealed $\langle vsini \rangle$ distributions of early- and late-type stars.

In many contemporary studies, $P{_{\rm rot}}$ is 
commonly used instead of direct measurements of stellar rotational velocities. 
However, in our dataset comprising 50,000 stars, our objective was
to demonstrate variations 
in stellar rotation across STs in a more transparent manner 
by utilising ${vsini}$. By focusing solely on single stars, we also sought to 
maintain the reliability of our analysis, with the expectation that this approach 
would enable more precise estimations of stellar velocities.
Moreover, our findings revealed a shift in the correlation between ST and
and $\langle vsini \rangle$ as stars evolve. 

In subsequent studies, 
the investigation will be expanded to encompass chemically peculiar stars 
and cluster member stars, with the objective of acquiring a more profound comprehension of the underlying mechanisms that regulate
the influence of stellar abundance on stellar rotation.

\section*{Acknowledgements}

In conducting this study, the bibliographic service NASA Astrophysics Data System, 
SIMBAD and VizieR online catalogue were employed.
This study was supported by the Scientific and Technological Research Council
of Turkey (TÜBİTAK) under Grant Number 2209A-1919B012310850. 
The authors thank TÜBİTAK for their support.

We would like to express our gratitude to İbrahim Küçük and Nur Filiz Ak 
for their valuable comments and insightful discussions on the manuscript.
Furthermore, we are indebted to Maria Granados Serrano and Sertaç Bera Yurtaslan 
for their assistance in reviewing and refining 
the language of the revised manuscript.

The {\sc{pygaus}} code, which is capable of performing 
averaging and Gaussian modelling of the 
targets, can be accessed at the following the URL:
https://github.com/MertBoranSt/PYGAUS.

\vspace{-1em}

\bibliography{bib} 


\appendix

\section{The overall $vsini$-ST distribution of the data set}

\begin{figure*}[!t]
\includegraphics[width=\textwidth]{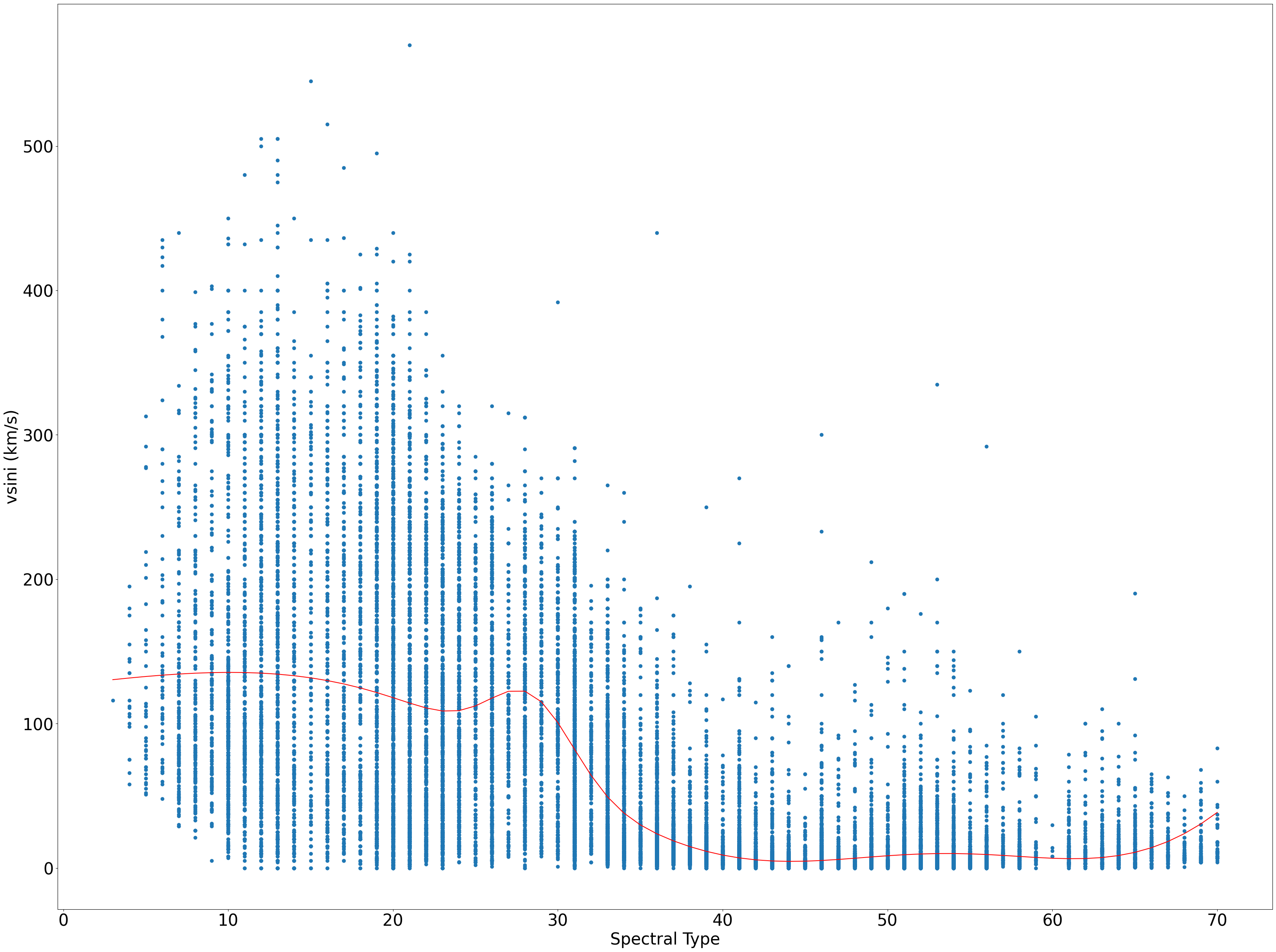}
\caption{The overall $vsini$-ST distribution of the data set. 
The x-axis and y-axis represent the STs, ranging from O0 to M9, and  
the $vsini$ values, respectively. The red line represents
 the Gaussian fit to all stars which are investigated in this study.}
\label{genel}
\end{figure*}

\section{{\sc{pygaus}} code}

The {\sc pygaus} code is utilised to ensure the efficient and consistent application of a Gaussian model to the pertinent data. The code incorporates Python libraries:
\texttt{matplotlib} \citep{Hunter2007}, \texttt{numpy} \citep{harris2020array}, 
\texttt{pandas}\footnote{https://pandas.pydata.org/} and
lmfit \citep{newville201511813}.

\end{document}